\documentclass[english,format=sigconf]{acmart}
\usepackage{babel}
\AtBeginDocument{%
  }

\setcopyright{none}
\acmDOI{}
\acmISBN{}

\acmConference[LOCO '26]{2nd Workshop on Low Carbon Computing}{Sept 10--11, 2026}{Lancaster and Online}

\usepackage[T1]{fontenc}
\usepackage[latin9]{inputenc}
\usepackage{varwidth}
\usepackage{graphicx}

\makeatletter

\providecommand{\tabularnewline}{\\}
\newenvironment{cellvarwidth}[1][t]
    {\begin{varwidth}[#1]{\linewidth}}
    {\@finalstrut\@arstrutbox\end{varwidth}}
\floatstyle{ruled}
\newfloat{algorithm}{tbp}{loa}
\providecommand{\algorithmname}{Algorithm}
\floatname{algorithm}{\protect\algorithmname}

\newenvironment{lyxcode}
	{\par\begin{list}{}{
		\setlength{\rightmargin}{\leftmargin}
		\setlength{\listparindent}{0pt}% needed for AMS classes
		\raggedright
		\setlength{\itemsep}{0pt}
		\setlength{\parsep}{0pt}
		\normalfont\ttfamily}%
	 \item[]}
	{\end{list}}

\makeatother
\usepackage{babel}
\begin{document}
\title{GPU implementation of a resource-constrained virtual machine}

\author{Simone Li}
\email{2827442L@student.glasgow.ac.uk}
\affiliation{%
  \institution{University of Glasgow}
  \city{Glasgow}
  \state{}
  \country{UK}
}
%\orcid{1234-5678-9012}

\author{Vladislav Brusokas}
\email{2840479B@student.glasgow.ac.uk}

\author{Andrei Ghita}
\email{2660223G@student.glasgow.ac.uk}

\author{Shuxuan Li}
\email{2782934L@student.gla.ac.uk}

\affiliation{%
  \institution{University of Glasgow}
  \city{Glasgow}
  \state{}
  \country{UK}
}

\author{Wim Vanderbauwhede}
\email{wim.vanderbauwhede@gla.ac.uk}

\affiliation{%
  \institution{University of Glasgow}
  \city{Glasgow}
  \state{}
  \country{UK}
}

\renewcommand{\shortauthors}{Li et al.}

\begin{abstract}

One of the main reasons compute hardware becomes obsolete is software bloat: resource requirements increase for every iteration of a software product. Resource constrained VMs are one way to combat software bloat as they post a hard limit on the resources and so force the programmer to be frugal. In this paper we explore the deployment of one such resource constrained VM, Uxn, on GPU. 

%Uxn is a severely resource constrained virtual machine (VM) with 64 kB memory and 8-bit instructions,
%embodying the philosophies of permacomputing and frugal computing.
%Implementations exist for many platforms and operating systems but
%were until now restricted to using the processor, with optional use of the GPU purely
%for graphics rendering and display. 
%
%We present an implementation of Uxn which performs all computation and graphics on the GPU. 

We show that for competitive performance it is essential to make use of the GPU data parallelism. We present an OpenMP-style parallelism API for Uxntal, the stack-based assembly-style language for the Uxn platform. We demonstrate that exemplar code using our API can run at comparable performance even on an integrated GPU. Specifically, our evaluation results show that using this approach
improves performance on the compute-intensive Stencil benchmark with
19$\times$ and frame rate on the graphics-intensive Bunnymark benchmark
with 7$\times$. 

In practice, all laptops and desktops and even mobile devices have a GPU and our work shows that they can be used to execute frugal workloads effectively. 
\end{abstract}
\maketitle

\section{Introduction}

Growing awareness of the contribution of emissions from computing
to climate change \cite{freitag2021real} has
sparked interest in ways to reduce emissions from computing.

Frugal computing advocates for treating
computational resources as finite and precious, to be utilised only
when necessary, and as frugally as possible \cite{vanderbauwhede2023frugal}.
It implies that computing needs to have the lowest possible energy
consumption and -- importantly -- use the currently available hardware
technologies. Since emissions from production of end-user hardware far exceed
the emissions from computing itself, extending the lifetimes of hardware
devices is important to curb emissions from computing.

Permacomputing, in analogy with permaculture, considers
that computing hardware should be designed for longevity
and software designed for resilience in case of (catastrophic) failure
of the supporting infrastructure such as the internet, hardware manufacturers,
software developing companies, etc. Software should also be understandable
and maintainable by a single person, which requires simplicity \cite{permacomputing2}.
%
%There has been interest in developing small resource-constrained virtual
%machines for some time. Examples include Ribbit, Contiki, Contiki-NG,
%TinyOS Mate \cite{10.1145/3486606.3486783,dunkels2004contiki,oikonomou2022contikiNG,levis2002mate},
%as well as various emulators of old hardware with a strong focus on
%games \cite{emulators,amiga,PCSX2}.

In this work we used the Uxn VM \cite{uxn}, developed by Hundred Rabbits
(100R), an artist collective that explores low-tech solutions. Uxn
is designed for sustainability and portability and is an 8-bit single-threaded
virtual CPU. Uxn memory is comprised of 64K of addressable memory
(RAM), two 256 byte stacks (return stack, working stack), and 256
bytes of device data. Uxn is programmable in Uxntal, a stack-based
assembly language. It is inspired by other stack-based languages (Forth
\cite{moore1970forth}) and by the assembly languages used in gaming
consoles. It defines its own graphical system and uses sprite-based
and pixel-based graphics to run small graphical applications. Uxn
supports up to 16 different input/output (I/O) devices; standard devices include
screen, audio, mouse, console, controller, file and datetime \cite{uxn}.
These devices have their own address space and the virtual machine
communicates with these through dedicated instructions. Events initiated
by the I/O devices are dealt with via event handlers ("vectors"
in Uxn).

Personal computing devices (desktops, laptops, phones, tablets) produced
within the last five to ten years all contain both a CPU and a GPU. GPUs
are powerful, massively parallel computational devices. The embodied
carbon of the GPU is as significant as that of the CPU, so if we wish
to use computing devices carbon-efficiently, we need to find ways
to make full use of the GPU. But GPUs are not designed
for general-purpose computing. The challenge we aim to address in our research is how to achieve
acceptable performance for programs running on a resource-constrained virtual machine implemented on a GPU
device. We have used the Uxn/Uxntal/Varavara
system to do so. The standard Uxn implementation by 100R uses the GPU for graphics rendering through the SDL2 library, however, all computation is performed entirely on the CPU.

%There is little prior art for VMs running on GPU and none for resource-constrained, graphics-capable ones such as Uxn.
% GVM \cite{celik2019design} is a GPU-Based Java Bytecode Interpreter. It can outperform a sequential execution on a CPU-based JVM interpreter and JVM with JIT, and produce comparable results to using JVM with JIT with many CPU threads. However, the GVM heavily leverages the parallelism provided by Java, whereas Uxn does not support parallelism. Furthermore, the workloads performed by GVM only operate with memory, which allows the work to be performed uninterrupted, whereas Uxn requires constant feedback and visual output, and thus needs to hand over control to the CPU constantly.

\section{Methodology}

We implemented Uxn on GPU using the Vulkan API for both compute and
graphics. 
%From experiments it emerged that optimisations were required
with regard to direct memory access, unifying shaders and concurrent event
handling. 
Uxn works well for lightweight graphical applications so
specific demanding applications were required for benchmarking. We
benchmarked the performance of the CPU and baseline GPU implementation,
with as metrics Frames per Second (FPS) and total runtime. We then
compared the performance of the optimised GPU implementation to the
baseline GPU implementation.

\subsection{GPU programming}

%The first iteration of implementing Uxn on GPU used OpenCL (Open Computing
%Language) and SDL (Simple DirectMedia Layer). OpenCL is an open, cross-platform
%standard for programming (amongst other devices) GPUs \cite{munshi2009opencl}.
%It is exclusively a compute framework and does not handle the graphical
%functionality of Uxn. This capability was implemented using SDL, an
%open, cross-platform standard for displaying graphics \cite{SDL}.
%However, due to the split between compute and graphical tasks, there
%were inefficiencies with this approach, which led to very low frame rates and overall poor performance.
%
%To overcome these issues, we subsequently 
We implemented Uxn using the
cross-platform Vulkan SDK which allows to combine both compute and
graphics. The SDK consists of a
C/C++ API for CPU programming and a domain-specific language, GLSL, for the code running on the GPU (known as "shaders"). The CPU-side API handles data movement between the host memory and the GPU memory and schedules the code to be run on the GPU.

\subsection{Uxn Implementation}

The aim of our work was to implement the entire Uxn virtual machine
and as much as possible of the I/O device functionality on the GPU,
thus minimising CPU-to-GPU transfer. However, this approach does not
entirely eliminate the need for transfer because the pipeline is CPU-controlled.

\subsubsection{Overall program structure}

\begin{figure}

\begin{centering}
\includegraphics[width=0.8\columnwidth]{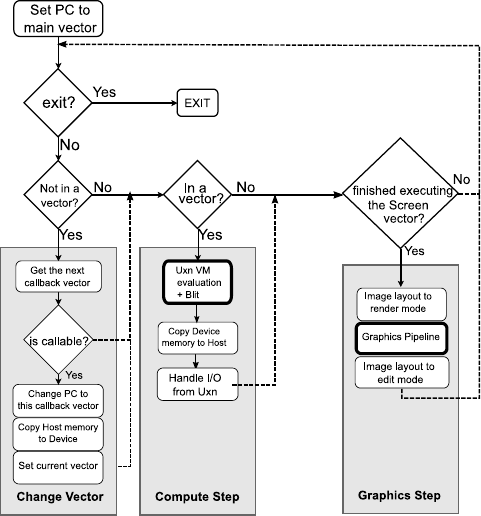}
\par\end{centering}
\caption{Flow chart showing the control flow and interactions between CPU and GPU. Blocks with a bold outline are executed on the GPU.}\label{fig:uxn-on-gpu-flowchart}
\end{figure}

Fig. \ref{fig:uxn-on-gpu-flowchart} shows the control flow of the host program and the interactions between CPU and GPU. The structure is complex as the code must deal both with I/O events generated on the host side (keyboard, mouse and the main clock event that governs the frame rate), and with I/O calls from the VM on the GPU (to write to the Console device and to draw on the screen). As mentioned above, Uxn is event-driven, so the I/O devices can register an event handler (called "vector") which will be triggered when an I/O event occurs. 
Therefore the main flow is (1) determine the next vector to be executed; (2) compute and render on the GPU; (3) draw to the screen using the GPU.  On every event that requires hand-over between the CPU and the GPU, the VM state must be copied to the host to ensure that it is preserved, as Vulkan does not guarantee that any state on the GPU is preserved between calls. 

\subsubsection{Uxn VM core on GPU}

We implemented the core of the Uxn VM, the actual bytecode interpreter which evaluates the instructions, using GLSL. The main loop of the VM is a direct port from the C code of the 100R implementation. The changes in this part relate mainly to the memory handling and the I/O device implementations, because the GPU can't handle I/O directly. Devices only either read data from the CPU or write data to the CPU. The exception is the Screen device, because drawing of pixels and sprites is part of the computation.

\subsection{Single-threaded performance}

Porting a single-threaded VM like Uxn to a GPU poses many performance
challenges. Fundamentally, whenever an I/O instruction is encountered,
the virtual machine must give control back to the CPU. This requires
the entire state to be copied back and forth. Furthermore, GPUs are
not optimised for control flow nor for calculations on small
integers. We optimised single-threaded performance by reducing host-to-device interaction and rewriting the VM interpreter to reduce the average number of conditions to be evaluated.

\subsection{Using GPU parallel threads}

\subsubsection{Rationale}

Single-threaded (``sequential'')
performance is slow compared to the 100R CPU version. This is
not surprising as GPU are designed for data parallelism. To exploit this without creating a complex
parallel programming API (such as e.g. the POSIX pthreads API) for
Uxntal, we opted for a very simple mechanism for parallelisation
of dependency-free loops, inspired by the OpenMP \cite{dagum1998openmp}
``parallel do'' pragma. All programmer needs to do is to tags a loop
and identify the loop bounds. This is necessary because Uxntal is
an assembly language and there are not syntactic constructs to indicate
loop structures. 

\subsubsection{Design and implementation}

Because of the limited functionality, the API is very minimal. We provide
 three Uxntal subroutines: 
\texttt{parallel-bounds}, 
\texttt{parallel-do} and
\texttt{end-parallel-do}. These functions are simple wrappers
around calls to a new I/O device that handles the parallelism (the
``Parallel'' device). The initial loop bounds must be on the stack, as is common practice in Uxntal. 

The \texttt{parallel-bounds} call returns the corresponding per-thread loop bounds. The \texttt{parallel-do} call instructs the ``Parallel'' device to enable use of parallel threads.
Code listing \ref{alg:Example-use-of} shows the parallelisation of
the core loop of one of the benchmarks (discussed in Section \ref{subsec:Benchmarks}).
The only change to the code is the addition of the three subroutine
calls (indented). On Uxn implementations that don't have
the Parallel device, these calls have no effect.

\begin{algorithm}
\begin{lyxcode}
;sprite/length~LDA2~\#0000

\textcolor{blue}{~~~~parallel-bounds~}

\textcolor{blue}{~~~~parallel-do}

\&loop~(~-{}-~)

EQU2k~?\&bail

DUP2~<draw-bunny>

INC2~!\&loop

\&bail~(~-{}-~)

POP2~POP2~

\textcolor{blue}{~~~~end-parallel-do}
\end{lyxcode}
\caption{Example use of the parallelisation API used in the Bunnymark benchmark }\label{alg:Example-use-of}

\end{algorithm}

Uxn communicates with external systems ("devices") via a 256-byte device page, providing 16 bytes of addressable space ("ports") for up to 16 devices. The Parallel device occupies one such slot. It uses one byte to indicate that the VM is in a parallel section (the \emph{control} port), and a short (2 bytes) for each of the loop bounds (the \emph{upper} and \emph{lower} ports).

Loop bounds are written to the respective ports, and setting the control byte marks the start of parallel execution. The device then distributes the loop iterations among invocations.

A fixed number of threads are dispatched at the start of shader execution. For the sequential part of the program, the main thread performs the evaluation while the worker threads stay idle. The worker threads operate on local stacks but on shared memory (the Uxn VM RAM). The local loop bounds and the thread IDs are pushed onto the per-thread stacks. Thread IDs allow thread-local state within the parallelised loop despite the shared RAM.

When the worker threads terminate, the main thread inherits the program counter of the worker thread with the highest ID and resumes the sequential execution.

\subsection{Benchmarks}\label{subsec:Benchmarks}

Uxn is predominantly made for lightweight graphical applications,
and the GPU version performs well on this type of applications, especially
those involving a lot of user interaction and not very demanding computation
or graphics. Therefore, to evaluate the performance relative to the
reference CPU implementation, we used specialised programs for benchmarking.

\begin{figure}

\begin{centering}
\includegraphics[width=0.4\columnwidth]{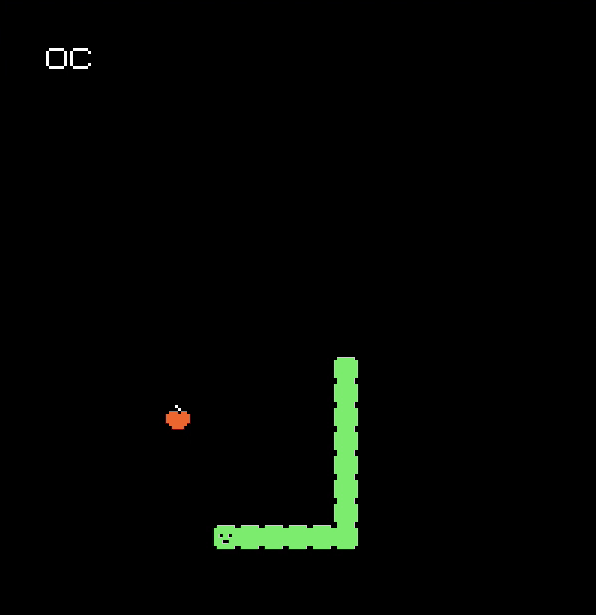}
\par\end{centering}
\caption{Screenshot of the Uxn version of the snake game in action}\label{fig:snake}
\end{figure}

An example of a lightweight graphical application where the baseline
GPU implementation worked very well is Snake \cite{snake}, implemented
in Uxntal in the \texttt{snake.tal} program \cite{uxntalsources}.
In this game the player controls a snake on a grid (Fig. \ref{fig:snake}). The aim is to
move the snake so it can eat fruits that appear randomly on the screen.
With each eaten fruit the snake's tail grows; however if the snake's
head collides with the snake's tail then the player loses. This game
is a good test for the performance of the overall virtual machine:
it requires keyboard or mouse support for controlling the snake as
well as support for the datetime device which is used for a random
number generator that determines the placement of the fruits.

To benchmark the performance of the CPU and baseline GPU implementation
and assess the performance of the optimised GPU implementation using
specialised programs, we used two main metrics: 
\begin{itemize}
\item Performance -- Frames per Second (FPS): the virtual machine should
be able to produce frames at a steady rate, even at high workloads. 
\item Performance -- Total Runtime: for programs with a set amount of workload,
the total runtime is used to compare performance. 
\end{itemize}
We used two specialised benchmark programs, Stencil and Bunnymark.

\subsubsection{Compute-only benchmark: Stencil}

The aim of the Stencil benchmark is to evaluate pure compute performance
without any graphics. The \texttt{stencil.tal} program implements
a 3-D stencil calculation as used in finite difference based differential
equation solvers, performed in a time loop. This
program is computationally and memory intensive, and provides clear
runtime duration for a set amount of work, comparable between implementations.
The pseudocode is shown in Listing \ref{alg:Stencil-calculation}

\begin{algorithm}
\begin{lyxcode}
do~t~=~1,te

p(i,j,k)~=~(

	(

	p(i+1,j,k)~+

	p(i-1,j,k)~+

	p(i,j+1,k)~+

	p(i,j-1,k)~+

	p(i,j,k+1)~+

	p(i,j,k-1)

	)/6~+

	p(i,j,k)

	)/2

end~do
\end{lyxcode}
\caption{Stencil calculation}\label{alg:Stencil-calculation}

\end{algorithm}

\subsubsection{Compute and graphics benchmark: Bunnymark}

Bunnymark \cite{bunnymark} is a popular testbench to assess raw
graphics performance. The \texttt{bunnymark.tal} program \cite{uxntalsources}allows
to create or remove jumping bunnies on the screen, displayed as sprites,
via key presses or mouse clicks. The path of the bunny is a parabola
reflecting the jumping movement and impact of gravity. For every frame,
for each bunny, calculations are performed to determine the next position
based on the current position and the direction of movement, and the
bunny is subsequently redrawn. The more bunnies there are, the higher
the load on the system, and the frame rate (frames per second, FPS)
is calculated and displayed by the program itself.

\begin{figure}
\begin{centering}
\includegraphics[width=0.8\columnwidth]{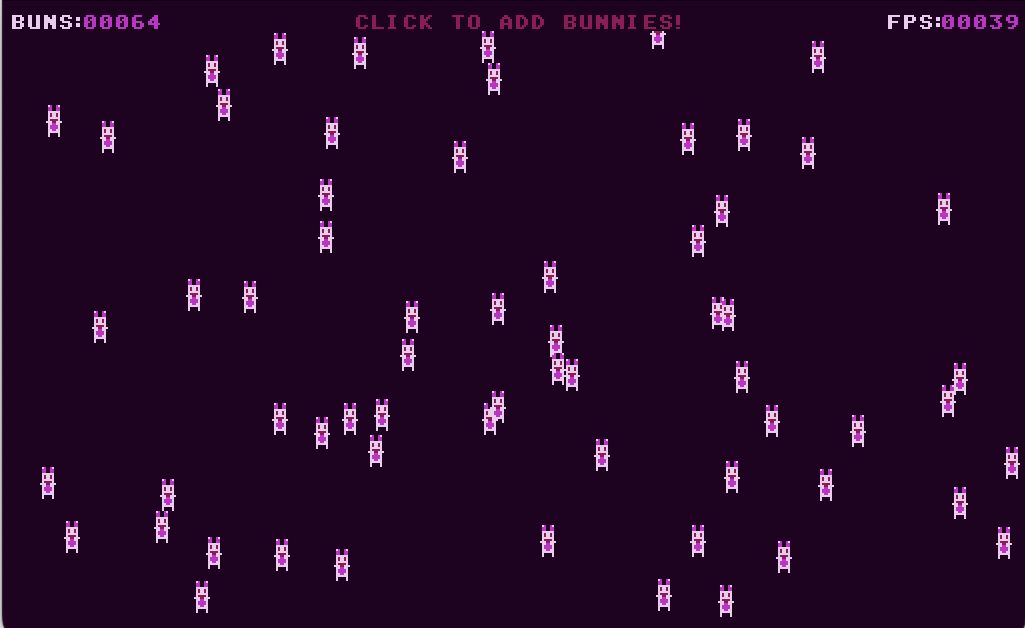}
\par\end{centering}
\caption{Screenshot of bunnymark in action}\label{fig:Screenshot-of-bunnymark}
\end{figure}

\subsection{Test platforms}

In keeping with the typical use of Uxn and the frugal computing philosophy,
the main test platforms used in the evaluation were two laptops from 2019.
Their specifications are shown in Table \ref{tab:Test-platforms-used}. 
%Additionally we use two more recent test platforms, shown in Table \ref{tab:Additional-test-platforms}. M1 is from 2021, AC from 2025.

\begin{table*}
\small
\caption{Test platforms used in the evaluation}\label{tab:Test-platforms-used}
\begin{tabular}{|c|c|c|c|c|c|}
\hline 
 & \multicolumn{3}{c|}{TongFang PF4MN2F (``TF'')} & \multicolumn{2}{c|}{MacBookPro16,1 (``MBP'')}\tabularnewline
\hline 
\hline 
OS & \multicolumn{3}{c|}{Ubuntu 22.04} & \multicolumn{2}{c|}{macOS 12.7.6}\tabularnewline
\hline 
 & Intel CPU & Intel GPU & NVIDIA GPU & Intel CPU & AMD GPU\tabularnewline
\hline 
Type & \begin{cellvarwidth}[t]
\centering
Core i7-10510U

(4 cores)
\end{cellvarwidth} & \begin{cellvarwidth}[t]
\centering
UHD Graphics 630

(24 CUs)
\end{cellvarwidth} & \begin{cellvarwidth}[t]
\centering
GeForce MX250

(3 CUs)
\end{cellvarwidth} & \begin{cellvarwidth}[t]
Core i7-9750H

(6 cores)
\end{cellvarwidth} & \begin{cellvarwidth}[t]
\centering
Radeon Pro 5500M

(22 CUs)
\end{cellvarwidth}\tabularnewline
\hline 
Clock speed (GHz) & 1.8--4.9 & 0.35--1.15 & 1.52--1.58 & 2.6--4.5 & 1.0--1.45\tabularnewline
\hline 
CMOS node & \multicolumn{2}{c|}{14 nm} & 14 nm & 14 nm & 7 nm\tabularnewline
\hline 
Die area (mm$^{2}$) & \multicolumn{2}{c|}{126 + 37} & 74 & 149 & 158\tabularnewline
\hline 
Thermal Design Power (W) & \multicolumn{2}{c|}{25} & 25 & 45 & 85\tabularnewline
%\hline 
% &  &  &  & \tabularnewline
\hline 
\end{tabular}
\end{table*}
%\begin{table*}
%\begin{tabular}{|c|c|c|c|c|}
%\hline 
% & \multicolumn{2}{c|}{acer PH18-73 (``AC'')} & \multicolumn{2}{c|}{MacBook Pro M1 (``M1'')}\tabularnewline
%\hline 
%\hline 
%OS & \multicolumn{2}{c|}{Windows 11} & \multicolumn{2}{c|}{macOS 26.3.2}\tabularnewline
%\hline 
% & CPU & GPU & \multicolumn{2}{c|}{CPU+GPU}\tabularnewline
%\hline 
%Type & \begin{cellvarwidth}[t]
%\centering
%Intel Core Ultra 9 275HX
%
%(16+8 cores)
%\end{cellvarwidth} & \begin{cellvarwidth}[t]
%\centering
%NVIDIA GeForce MX250
%
%(84 CUs)
%\end{cellvarwidth} & (8 cores) & (8 CUs)\tabularnewline
%\hline 
%Clock speed (GHz) & 2.7--4.9 & 2.3--2.6 & 2.0--3.2 & --\tabularnewline
%\hline 
%CMOS node & 3 nm & 5 nm & \multicolumn{2}{c|}{5nm}\tabularnewline
%\hline 
%Die area (mm$^{2}$) & 135+23+100+28 & 378 & \multicolumn{2}{c|}{120}\tabularnewline
%\hline 
%Thermal Design Power (W) & 55 & 360 & \multicolumn{2}{c|}{61}\tabularnewline
%\hline 
%\end{tabular}\caption{Test platforms used for additional results}\label{tab:Additional-test-platforms}
%\end{table*}

The clock speed lists two values. The highest value is the boost or turbo boost frequency. 
%All platforms except the M1 also include an Intel UHD Graphics integrated GPU, but the current implementation does not support those devices. 
Both platforms also include an Intel UHD Graphics integrated GPU, but the current implementation only supports this device on the TF. On both platforms we used the Vulkan SDK v1.4 and gcc/g++ version 13. 
%except on AC, where it was Microsoft Visual C++ for Windows 11. To assemble the Uxntal programs into ROMs, we use the 100R implementation of the Uxntal assembler.

\section{Evaluation}

We compared the performance of the GPU implementation against the
100R implementation of Uxn, both for sequential and parallel versions
of the benchmarks.

\subsection{Stencil}

A summary of the Stencil benchmark results is shown in Table \ref{tab:Summary-of-Stencil}.
The key observation is that the parallel version is almost twenty times faster
than the sequential version on all platforms, despite the very
different GPUs. On the integrated GPU on TF, this is still 4$\times$ slower than the reference;
on the discrete GPU on MBP, it is only $3\times$ slower. The difference is caused by the
better performance of the 100R Uxn VM implementation on the TF CPU
compared to the MBP CPU. This is interesting as the TF CPU has a much
lower base clock speed (1.8 GHZ vs 2.6 GHz). 

\begin{table}
\small
\caption{Summary of Stencil benchmark results, 16 iterations. \textquotedblleft Seq\textquotedblright{}
= sequential version on GPU; \textquotedblleft Par\textquotedblright{}
= parallel version on GPU; Ref = reference version on CPU }\label{tab:Summary-of-Stencil}
\begin{tabular}{|c|c|c|c|c|c|}
\hline 
\begin{cellvarwidth}[t]
\centering
GPU
\end{cellvarwidth} & \begin{cellvarwidth}[t]
\centering
Seq/Ref 

slowdown
\end{cellvarwidth} & \begin{cellvarwidth}[t]
\centering
Par/Seq 

speedup
\end{cellvarwidth} & \begin{cellvarwidth}[t]
\centering
Par/Ref 

slowdown
\end{cellvarwidth} & \begin{cellvarwidth}[t]
\centering
Par 

Elapsed 

time (s)
\end{cellvarwidth} & \begin{cellvarwidth}[t]
\centering
Ref 

Elapsed 

time (s)
\end{cellvarwidth}\tabularnewline
\hline 
\hline 
TF NVIDIA& $95\times$ & $19\times$ & $5\times$ & 1.23 & 0.24\tabularnewline
\hline 
TF Intel & $69\times$ & $18\times$ & $4\times$ & 0.92 & 0.24\tabularnewline
\hline 

MBP AMD & $53\times$ & $19\times$ & $3\times$ & 1.26 & 0.44\tabularnewline
\hline 
\end{tabular}

\end{table}

\begin{figure}

\includegraphics[width=1.0\columnwidth]{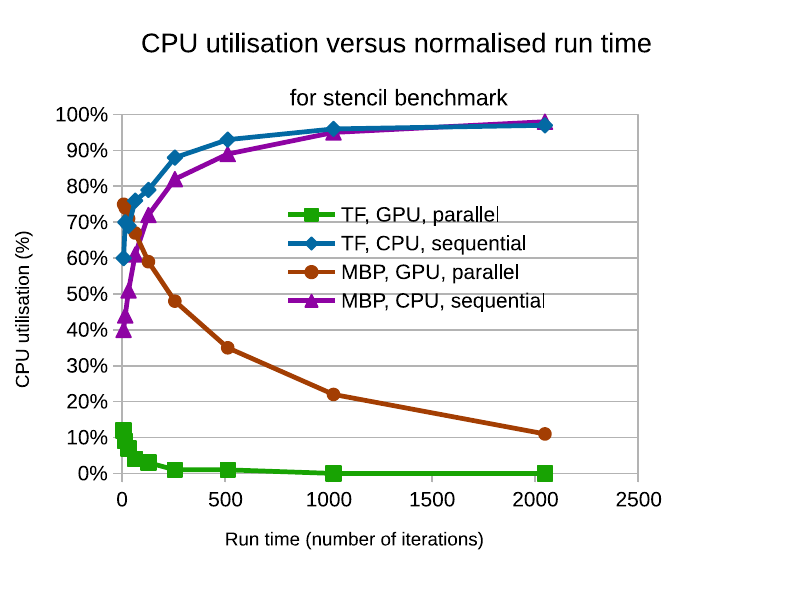}

\caption{CPU utilisation for the benchmark running on the GPU and on the CPU,
as a function of the number of iterations}\label{fig:CPU-utilisation-for}

\end{figure}

Another interesting observation is shown in Fig. \ref{fig:CPU-utilisation-for}.
It shows that for long running tasks, the CPU utilisation for the
reference implementation tends to 100\%, but for the GPU implementaton
it tends to 0\%. This confirms that the workload is indeed entirely
offloaded to the GPU. The difference in the utilisation between TF
and MBP is likely caused by the difference in CPU scheduling, and
therefore CPU time accounting, between the Darwin OS kernel of macOS
and the Linux kernel.

\subsection{Bunnymark}

\begin{figure}
\includegraphics[width=0.8\columnwidth]{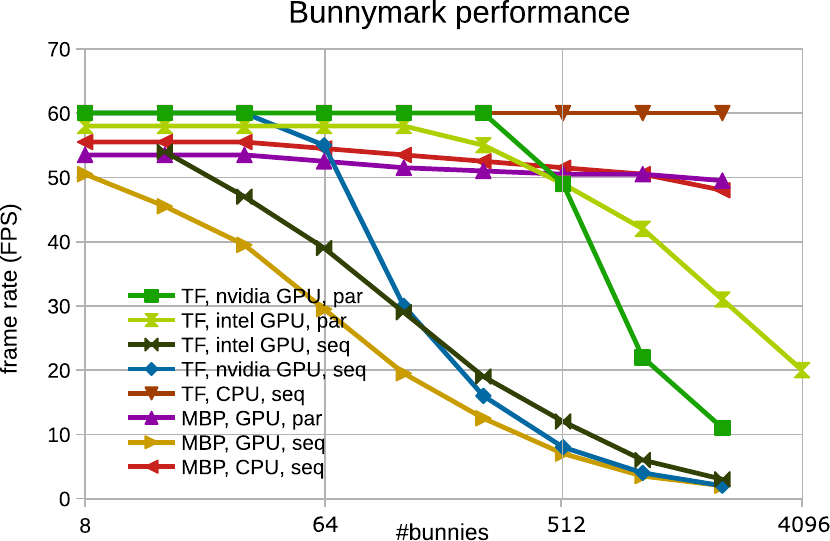}

\caption{Bunnymark performance }\label{fig:Bunnymark-performance}

\end{figure}

Bunnymark assesses the performance of the frame rate (FPS) as a function
of the number of bunnies. Fig. \ref{fig:Bunnymark-performance} shows
the GPU sequential and parallel performance and the performance of
the reference implementation. On the integrated GPU on TF, the optimised sequential version manages an acceptable 40 FPS
for 32 bunnies. The parallel version on TF tracks the reference version
until 256 bunnies and performs very well (>30FPS) up to 2048 bunnies. On MBP, the optimised
sequential version also manages 40 FPS for 32 bunnies. Performance
of the parallel version is on par with the reference version, maintaining
50 FPS for up to 2048 bunnies.  The performance of the sequential programs on the GPU implementation is in line with our earlier observation that most Uxn applications run fine on the sequential version because they are not very demanding. Parallelisation results in an improvement of 8$\times$ in FPS @ 512 bunnies for the integrated GPU on TF and 7$\times$ for the discrete GPU on MBP. 
% For the additional platforms, on M1, the parallel version performance is on par with the reference, and on AC, which has a very powerful GPU, it even outperforms it.

%
%\begin{table}
%\begin{centering}
%\begin{tabular}{|c|c|c|c|}
%\hline 
%\multicolumn{4}{|c|}{M1}\tabularnewline
%\hline 
%\#Bunnies & Seq & Par & Ref\tabularnewline
%\hline 
%15 & 43 & 48 & 41\tabularnewline
%\hline 
%127 & 13 & 48 & 40\tabularnewline
%\hline 
%4095 & 1 & 36 & 37\tabularnewline
%\hline 
%\hline 
%\multicolumn{4}{|c|}{AC}\tabularnewline
%\hline 
%\#Bunnies & Seq & Par & Ref\tabularnewline
%\hline 
%17 & 59 & 59 & 54\tabularnewline
%\hline 
%127 & 19 & 59 & 52\tabularnewline
%\hline 
%4095 & 1 & 59 & 49\tabularnewline
%\hline 
%\end{tabular}
%\par\end{centering}
%\caption{Additional Bunnymark performance figures on M1 and AC}
%
%\end{table}

\section{Discussion}

Many aspects conspire to make the current implementation of the Uxn VM slow on the GPU: the VM needs frequent data movement and control hand-over between CPU and GPU because of I/O handling; the interpreter is single-threaded and control flow dominated; and the VM compute consists entirely of integer operations. The compute-only benchmark shows that pure compute using a single thread is 50$\times$ to 100$\times$ slower on GPU than on CPU. 
%We confirmed via micro-benchmarks that this is not a specific issue with the Uxn VM: it is a fundamental issue with the type of operations the VM needs. In particular, control flow (if-then or switch-case) in a tight loop is 30$\times$ slower on the MBP AMD GPU and 30$\times$ on the TF NVIDIA GPU. 
%As already discussed, 
Data movement also causes slowdown because we need to copy memory between the host and the GPU memory, whereas the entire state of the Uxn VM on CPU fits easily in the cache. Furthermore, the GPU runs at a lower clock speed than the CPU. On TF, the observed CPU clock speed while running the reference implementation is 2.3 GHz; the observed GPU clock speed while running the integrated GPU implementation is 1.13 GHz, so a ratio of 2. On the MBP, the CPU clock runs at 2.2 GHz. The AMD GPU has a clock speed of at most 1.45 GHz so the ratio is around 1.5. All these aspects combine to result in very slow single-threaded performance for heavy workloads. As shown above, exposing the GPU data parallelism to the programmer results in considerable improvement in performance, both for compute and graphics, across all tested platforms. 

Using diagnostic tools (powertop, intel-gpu-top and nvidia-smi on TF, powermetrics on MBP) we have ascertained that the parallelism does not result in appreciably higher power consumption. This is because, although the GPUs have hundreds to thousands of hardware threads, they are grouped in far fewer compute units (e.g. 3 on TF, 24 on MBP), and therefore even a single thread fully occupies such a unit. For graphical applications, the aim is to compute in less than a frame period. If the computation is performed in parallel, it finishes sooner, so the energy consumption is reduced proportionally. On TF integrated GPU, there is a considerable energy saving from parallelisation as we speed up the computation 18$\times$ using parallelism at no observed increase in power consumption. We have further ascertained that running the Uxn VM on the GPU has a comparable power consumption to running it on the CPU (between 1$\times$ and 2$\times$). This may seem slightly odd because running a single-threaded VM on a multicore CPU should result in only a single core being occupied, and therefore the total power consumption should be proportionally lower. However, the idle power consumption is still considerable, and although the GPU spends more time doing the same work, the GPU implementation does not actually execute considerably more instructions: the main additional work is moving the data between the host and GPU memory, and the state of the Uxn VM is a very small amount of data. Therefore the power consumption is comparable. So overall, our results show that it is possible to obtain comparable performance at comparable power consumption for applications that can be parallelised, although for most cases the performance is slightly lower and the power consumption slightly higher.

It should be noted that in any case, the energy consumption of Uxn applications is low because for a typical interactive application with simple graphics, most of the time the process will be waiting for I/O operations as human input is very slow compared to the clock speed. Therefore, if we consider a laptop as a platform for running Uxn, rather than e.g. to watch streaming video, the environmental impact is mostly caused by its production, and in particular by the chip manufacturing process. %The greenhouse gas emissions caused by this process are called embodied or embedded carbon. 
According to a report from 2019 by the European Environmental Bureau (EEB) \cite{coolproducts}, for a laptop running average work workloads, ``manufacturing, distribution and disposal account for about 52\% of a notebook's total climate impact''. 

%Meanwhile, the carbon intensity of electricity generation (CI) of the EU has dropped from 253 gCO2e/kWh in 2019 to 187 gCO2e/kWh in 2024 \cite{eu_ci}, but the CI of Taiwan, where most of the chips are produced, has decreased only by 2\% \cite{owid_taiwan_ci}. The emission from the electricity used in the chip manufacturing process are only on a part of the embodied carbon of the chips: the materials used also contribute significantly. As a result, the contribution of the embodied carbon has proportionally 
This proportion has 
risen to 65\% in 2025. The EEB calculates that to limit the Global Warming Potential, we should use our laptops for 20 to 44 years, whereas the current average lifetime is 4.5 years. Making better use of the available hardware is one way to achieve this, and that is our main motivation to make Uxn run on GPUs, not because of Uxn specifically but because resource constrained VMs are one way to combat the software bloat which is one of the main reasons  compute hardware becomes obsolete.

\section{Conclusion}

We have presented the GPU implementation of a resource-constrained virtual machine, the Uxn VM, and a simple OpenMP-style parallelism API for Uxntal, the stack-based assembly language for this platform. Even without making use of the GPU hardware parallelism, the GPU implementation can run most Uxn programs with the same performance as the reference CPU implementation, but we have found that single-threaded GPU performance is intrinsically much worse than that of the CPU. To make optimal used of the GPU, using data-parallel computation is essential. We have shown that our simple approach to parallelise dependency-free loops results in very considerable performance improvements without increasing power consumption. On all test platforms, the resulting performance on the Bunnymark benchmark was comparable with the CPU implementation, with comparable power consumption. 

In future work we will further optimise the performance and investigate if there are specific use cases where Uxn on the GPU outperforms the CPU.
%further optimise the Uxn VM for the GPU, in particular redesigning the control flow of the VM and investigating improved I/O buffering. 
The source code for this work has been released under an open source license and can be found at \cite{uxn_on_gpu}.
\bibliographystyle{unsrtnat} 
\bibliography{paper-LOCO2026-uxn-on-gpu}

@article{freitag2021real,
  title={The real climate and transformative impact of ICT: A critique of estimates, trends, and regulations},
  author={Freitag, Charlotte and Berners-Lee, Mike and Widdicks, Kelly and Knowles, Bran and Blair, Gordon S and Friday, Adrian},
  journal={Patterns},
  volume={2},
  number={9},
  year={2021},
  publisher={Elsevier}
}

@misc{permacomputing2,
  title={Permacomputing Update 2021},
  author={Heikkil\"a, V.-M.},
  booktitle={},
  year={2021},
  month={},
  publisher={},
  url={http://viznut.fi/texts-en/permacomputing_update_2021.html.},
  urldate={2026-03-18}
}

@article{vanderbauwhede2023frugal,
  title={Frugal Computing--On the need for low-carbon and sustainable computing and the path towards zero-carbon computing},
  author={Vanderbauwhede, Wim},
  journal={arXiv preprint arXiv:2303.06642},
  year={2023}
}

@misc{uxn,
  title={Hundred rabbits},
  author={Lu Linvega, D.},
  booktitle={},
  year={},
  month={},
  publisher={},
  url={https://wiki.xxiivv.com/site/hundred_rabbits.html},
  urldate={2026-03-18}
}

@article{moore1970forth,
  title={Forth--a language for interactive computing},
  author={Moore, Charles H and Leach, Geoffrey C},
  journal={Amsterdam: Mohasco Industries Inc},
  year={1970}
}

@misc{uxntalsources,
  title={Hundred rabbits},
  author={Lu Linvega, D.},
  booktitle={},
  year={},
  month={},
  publisher={},
  url={https://git.sr.ht/~rabbits/uxn-games/},
  urldate={2026-03-18}
}

@book{stencil,
	author = {Milne, William Edmund},
	title = {Numerical Solution of Differential Equations.},
	publisher = {Dover Publications},
	year = {1970},
	address = {New York,},
	edition = {2d rev. and enl. ed.}
}

@misc{bunnymark,
    title = {Display list vs. blitting - the results!
},
    author = {Lobb,Iain},
    url={https://blog.iainlobb.com/2010/11/display-list-vs-blitting-results.html},
    urldate={2026-03-18},
    year={2010},
    month={nov}
}

@misc{snake,
    author= {Quinn Myers},
    title={An Oral History of Snake on Nokia},
booktitle={MEL Magazine},
year={2019},
url={https://melmagazine.com/en-us/story/snake-nokia-6110-oral-history-taneli-armanto},
 urldate={2026-03-18},
}

@article{dagum1998openmp,
  title={OpenMP: an industry standard API for shared-memory programming},
  author={Dagum, Leonardo and Menon, Ramesh},
  journal={IEEE computational science and engineering},
  volume={5},
  number={1},
  pages={46--55},
  year={1998},
  publisher={IEEE}
}

@misc{coolproducts,
author={EEB},
title={Coolproducts don't cost the earth -- full report},
year={2019},
url={www.eeb.org/coolproducts-report},
 urldate={2026-03-18},
}

@misc{uxn_on_gpu,
title = {{Uxn on GPU}},
author = {Li, Simone and Brusokas, Vladislav Ghita, Andrei and Li, Shuxuan and Vanderbauwhede, Wim
},
url={https://codeberg.org/wimvanderbauwhede/uxn-gpu},
urldate={2026-03-26}
}

\end{document}